\documentclass{elsart}
\usepackage{epsfig}
\usepackage{color}

\makeatletter
\newdimen\z@ \z@=0pt 
\newskip\z@skip \z@skip=0pt plus0pt minus0pt
\def\m@th{\mathsurround=\z@}
\def\ialign{\everycr{}\tabskip\z@skip\halign} 
\def\eqalign#1{\null\,\vcenter{\openup\jot\m@th
  \ialign{\strut\hfil$\displaystyle{##}$&$\displaystyle{{}##}$\hfil
      \crcr#1\crcr}}\,}
\makeatother

\newcommand{\aff}[2]{Dipartimento di Fisica dell'Universit\`a #1 e 
Sezione INFN, #2, Italy.}
\newcommand{\affd}[1]{Dipartimento di Fisica dell'Universit\`a e Sezione 
INFN, #1, Italy.}

\newcommand{\eeg}{$e^+ e^- \gamma$}
\newcommand{\mmg}{$\mu^+ \mu^- \gamma$}

\newcommand{\ppp}{$\pi^+ \pi^- \pi^0$}
\newcommand{\mtrk}{$m_{\rm trk}$}
\newcommand{\amu}{$a_\mu$}
\renewcommand{\to}{\ensuremath{\rightarrow}}
\renewcommand{\L}{\ensuremath{{\mathcal L}}}

\def\ifm#1{\relax\ifmmode#1\else$#1$\fi}
\def\sig{\ifm{\sigma}} \def\epm{\ifm{e^+e^-}}

\def\pic{\ifm{\pi^+\pi^-}} \def\gam{\ifm{\gamma}} 
\def\po{\ifm{\pi^0}}
\def\up#1{$^{#1}$}  \def\dn#1{$_{#1}$}  \def\DAF{DA\char8NE}
\def\ab{\ifm{\sim}}  \def\x{\ifm{\times}}  \def\ff{$\phi$--factory}  
\def\ks{\ifm{K_S}} \def\kl{\ifm{K_L}} 
\def\ie{{\it\kern-2pt i.\kern-.5pt e.\kern-2pt}}  \def\f{\ifm{\phi}}
\def\dif{\hbox{d\kern.1mm}}  \def\deg{\ifm{^\circ}}
\def\up#1{$^{#1}$}  \def\dn#1{$_{#1}$}
\def\pt#1,#2,{\ifm{#1\x10^{#2}}}
 
\def\bye{\end{document}}

\begin{document}

\begin{flushright}
    KLOE Note n$^o$ 189

    July 2003
\end{flushright}
\begin{frontmatter}


\title{Determination of $\sigma(e^+e^-\rightarrow \pi^+ \pi^-)$ from radiative processes at \DAF}

\vskip -1cm

\collab{The KLOE Collaboration}

\author[Na] {A.~Aloisio},
\author[Na]{F.~Ambrosino},
\author[Frascati]{A.~Antonelli},
\author[Frascati]{M.~Antonelli},
\author[Roma3]{C.~Bacci},
\author[Frascati]{G.~Bencivenni},
\author[Frascati]{S.~Bertolucci},
\author[Roma1]{C.~Bini},
\author[Frascati]{C.~Bloise},
\author[Roma1]{V.~Bocci}
% ...

\clearpage
\address[Bari]{\affd{Bari}}
\address[Frascati]{Laboratori Nazionali di Frascati dell'INFN,
Frascati, Italy.}
\address[Karlsruhe]{Institut f\"ur Experimentelle Kernphysik,
Universit\"at Karlsruhe, Germany.}
\address[Lecce]{\affd{Lecce}}
\address[Na]{Dipartimento di Scienze Fisiche dell'Universit\`a
``Federico II'' e Sezione INFN,
Napoli, Italy}
\address[Pisa]{\affd{Pisa}}
\address[Energ]{Dipartimento di Energetica dell'Universit\`a
``La Sapienza'', Roma, Italy.}
\address[Roma1]{\aff{``La Sapienza''}{Roma}}
\address[Roma2]{\aff{``Tor Vergata''}{Roma}}
\address[Roma3]{\aff{``Roma Tre''}{Roma}}
\address[StonyBrook]{Physics Department, State University of New
York at Stony Brook, USA.}
\address[Trieste]{\affd{Trieste}}
\address[Virginia]{Physics Department, University of Virginia, USA.}
\address[Beijing]{Permanent address: Institute of High Energy
Physics, CAS,  Beijing, China.}
\address[Moscow]{Permanent address: Institute for Theoretical
and Experimental Physics, Moscow, Russia.}
\address[Tbilisi]{Permanent address: High Energy Physics Institute, Tbilisi
  State University, Tbilisi, Georgia.}

\begin{abstract}
We have measured the cross section \sig(\epm\to\pic\gam) with the KLOE detector at \DAF, at an energy $W=M_\phi=1.02$ GeV. From the dependence of the cross section on $m(\pic)=\sqrt{W^2-2WE_\gamma}$, where  $E_\gamma$ is the energy of the photon radiated from the initial state, we extract \sig(\epm\to\pic) for the mass range $0.35<m^2(\pi^+\pi^-)<0.95$ GeV$^2$. From our result we extract the pion form factor and the hadronic contribution to the muon anomaly, $a_\mu$.
\end{abstract}
\end{frontmatter}

\section{Hadronic Cross Section at DA$\Phi$NE}

\subsection{Motivation}

The recent precision measurement of the muon anomaly \amu\ 
at the Brookhaven National Laboratory~\cite{e821}
has led to renewed interest in accurate measurements of the cross
section for $e^+e^-$ annihilation into hadrons. Contributions 
to the photon spectral functions due to quark loops, 
are not calculable for low hadronic mass states because of the
failure of perturbative QCD in such conditions. A very 
clean way out has been known for a long time~\cite{durand}. The 
imaginary part of the hadronic piece of the spectral function 
is connected by unitarity to the cross section for \epm\to hadrons. A dispersion relation can thus be derived,
giving the contribution to \amu\ as an integral over the hadronic 
cross section multiplied by an appropriate kernel. 
An example of a first 
complete estimate of the correction was given in 1985, 
$\delta a_\mu^{\rm had} $=\pt707(6)(17),-10, \cite{kinoshita}.
The process \epm\to\pic\ contributes \ab500 out of the \ab700 
value above. The cross section for \epm\to\pic\ becomes negligible 
above 1 GeV.

The most recent measurements of \sig(\epm\to\pic) in the energy interval 
610$<m(\pic)<$961 MeV come from CMD-2 at Novosibirsk. They claim a systematic error of 0.6\% with a statistical error of \ab0.7\% \cite{cmd2}.
Their data have been used most recently together with $\tau$ and \epm\ 
data up to 3 GeV, in an attempt to produce a firm prediction for 
comparison with the BNL result\cite{davhoe}. 
There is unfortunately a rather strong disagreement 
between the $\delta a_\mu^{\rm had}$ value obtained using 
\epm\to\pic\ data and $\tau$ decay data, after isospin corrections. 
Finally the \epm\to\pic\ based result disagrees by \ab3 \sig\ 
with the BNL \cite{e821} measurements. There are thus many reasons for new measurements of the \epm\ annihilation cross section into two pions.

\subsection{Initial State Radiation}

Initial state radiation, ISR, is a convenient mechanism which allows studying \epm\to\ hadrons, over the entire energy range from 2$m_\pi$ to $W$, the center of mass energy of the collision. 
In the case of interest it is potentially vitiated by the 
possibility of final state radiation. For a photon radiated 
before annihilation of the \epm\ pair, the \pic\ system energy is $m(\pic)=\sqrt{W^2-2WE_\gam}$, thus one measures the coupling to pions 
of an off-mass shell photon of mass m(\pic). For a photon radiated 
by the final state pions, the virtual photon coupling to the \pic\ 
pair has a mass $W$. By just counting powers of $\alpha$, the relative probability of ISR and FSR are equal. This requires very careful estimates of the two processes in order to be able to use the reaction \epm\to\pic\gam\ to extract \sig(\epm\to\pic). The Karlsruhe group has given much attention 
to this problem and has also developed the EVA and PHOKHARA Monte Carlo 
programs \cite{binner} - \cite{venanzo} which are fundamental to our analysis. 
In the following we will refer to the invariant mass squared of the \pic\ system as $s_\pi$. In general the \pic\gam\ and \pic\ cross section are related through:
$$s_\pi\,{\dif\sig(\pic\gam)\over\dif s_\pi\,\dif\cos\theta}=\sig(\pic,\;s_\pi)\x F(s_\pi,\;\theta_\gam)$$
Integrating the cross section above, for $0<\theta_\gamma<\overline\theta$ and $180\deg-\overline\theta<\theta_\gamma<180$ we get
\begin{equation}
s_\pi\,\frac{\dif\sigma(\pic\gam)}{\dif s_\pi}=
\sigma(\pic,\;s_\pi)\x H(s_\pi,\;\overline\theta)
\label{eq:H}
\end{equation}
Eq. \ref{eq:H} defines the radiator function $H(s_\pi,\;\overline\theta)$. In the following we will drop the variable $\overline\theta$ which is a constant in the present work. The radiator function $H$ used in our analysis is obtained from the PHOKHARA Monte Carlo program.
Our present analysis is based on the observation  of 
ref.\cite{binner}, that for small polar angle of the radiated photon, 
the ISR process vastly dominates over the FSR process. 
At lowest order the \pic\gam\ cross section diverges as 
$1/\sin^2\theta$, just like \sig(\epm\to\gam\gam). 
We limit ourselves in the following to studying the 
reaction \epm\to\pic\gam\ with $\theta_\gamma<15\deg$ or 
$\theta_\gamma>165\deg$. For small $m(\pic)$, the di-pion 
system recoiling against a small angle photon will results in 
one or both pions being lost also at small angle. 
We are therefore limited to measuring \sig(\pic) for $m(\pic)>$550 MeV. 
We will be able to investigate the cross section near threshold, as 
soon as next to leading order calculation for FSR become available. 
This is of great importance, since there are no recent, 
good measurements of \sig(\pic) at low mass, 
which weigh strongly in the estimate of $\delta a_\mu^{\rm had}$.

\section{The KLOE detector}
The KLOE detector \cite{kloe} consists of only two elements: a large, precision drift chamber and an electromagnetic calorimeter. The detector design was optimized for $CP$ study in neutral kaon decays, for K mesons produced in the decay of $\phi$ mesons almost at rest, $p_\phi$\ab12.5 MeV/c.
The measurements reported in the following rely heavily on the drift chamber. Its large dimensions, 4 m diameter, use of He and a magnetic field of 0.52 T allow measuring momenta and direction of charged particles with very high accuracy.
Because of our choice of accepting only events with $|\cos\theta_\gamma|>\cos15\deg$, the photon does not reach the calorimeter. The two relevant parameters, $s_\pi$ and $\theta_\gamma$ are obtained from the reconstructed tracks of events with two opposite charge particles forming a vertex near the \epm\ interaction point. The quality of the detector is fully described by the mass squared resolution and the resolution in measuring the production angle of single tracks or the angle of the vector sum of the two track momenta. We find $\sigma(m(\pic))/m$\ab0.2\% for $m$=500 MeV, without much change over the range studied and $\sigma(\theta_\gamma)$\ab10 mrad, with $\theta_\gamma=\cos^{-1}({\bf p}_{+-}\cdot\hat z)$ and ${\bf p}_{+-}=-({\bf p}^++{\bf p}^-)/|({\bf p}^++{\bf p}^-)|$. The momentum resolution is $\sigma(p_\perp)/p_\perp\leq0.4\%$ for 
large angle tracks. The calorimeter still plays an important role in the measurement. Its exceptional timing resolution, $\sig(t)=57/\sqrt{E\hbox{ GeV}}\oplus50$ ps and high segmentation provide particle identification allowing us to distinguish pion from electrons and muons, as discussed later. It is also used in the luminosity measurement,

The KLOE detector is operated at \DAF, the Frascati \ff, running at $W=m_\phi=1019.4$ MeV. Beams cross in \DAF\ at an angle of 25 mrad, resulting in a momentum of the $\phi$ mesons of \ab12.5 MeV/c along the $x$-axis. We use a coordinate system where the $x$-axis points to the center of the colliders, the $z$-axis bisects the two beam lines and the $y$-axis is vertical. The precise values of $W$ and $p_\phi$ is measured on a run by run basis using Bhabha events, to accuracies significantly better than 100 keV. The rms machine energy spread is \ab350 keV, measured with \ks\kl\ pairs in KLOE.

\section{Event Selection}
The excellent KLOE resolution allows us to efficiently selected \pic\gam\ events, without photon detection, because of their clean signature: two tracks from the interaction point. We assume that the event is of the type \epm\to$x^+x^-$\gam, with $m(x^+)=m(x^-)$ and $m_x=m_\pi$. The undetected photon polar angle, $\theta_\gam$ is the angle with the $z$-axis of the vector -{\bf p}\dn+-{\bf p}\dn-.

\subsection{Fiducial Volume}

The fiducial volume for \pic\gam\ events has been chosen in order to maximize the acceptance for \pic\gam\ events due to ISR. The main background processes are \epm\to\f\to\pic\po\ and radiative Bhabha scattering. At low mass the final state $\mu^+\mu^-\gam$ becomes important  but requiring small $\theta_\gamma$ removes most of this final state. We rely otherwise on kinematics and particle identification, ID, in order to reject the mentioned backgrounds to a negligible level. From studies with EVA, confirmed with PHOKHARA, a good compromise for the fiducial volume is $|\cos\theta_\gamma|>\cos15\deg$ and $50\deg<\theta_{+-}<130$ for the polar angle of both tracks at the interaction point. These requirements ensure abundant statistics for the measurement and contamination of FSR hard radiation well below the 1\% level. The efficiency for the events falling inside the fiducial cuts is contained in the value of the radiator, obtained from PHOKHARA.  We note that no detected photon is required, nor are events rejected if photons are present.

\subsection{Event Selection}
\subsubsection{Event origin} The two tracks must form a vertex close to the interaction point. We require that the vertex coordinates $x_V,x_y, x_z$ satisfy $\sqrt{x_V^2 + y_V^2}\le8$ cm and $|z_V|\le7$ cm. 
We also require $P_\perp>160$ or $P_z>90$ MeV/c to remove spiraling tracks.
90 to 95\% of the \pic\gam\ events survive these requirements. This result is obtained from Monte Carlo studies and examination of \f\to\pic\ decays.
\subsubsection{Pion identification} 
\begin{figure}[ht]
\begin{center}
\mbox{\epsfig{file=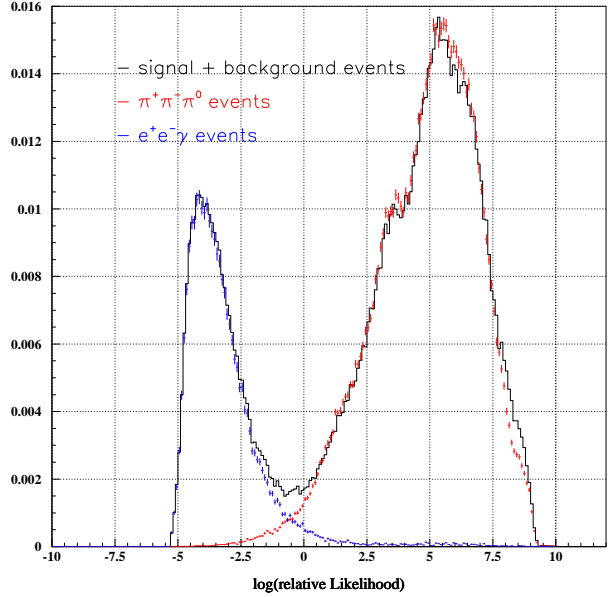,width=8cm}}
\caption{Distribution of the particle ID variable $\zeta$ for pions, right peak, and electrons, left. See text.}
\label{likelihood}
\end{center}
\end{figure}

A fraction of radiative Bhabha events, \epm\gam, survives kinematical selection, see sec. \ref{sec:kincut} below, contributing a non-negligible background. In order to reject these 
events, we have developed a particle ID method, using approximate likelihood estimators which are effective in distinguishing pion and electrons. The likelihood is based on time-of-flight vs momentum comparison and on the shape and magnitude of the energy deposits in the calorimeter by the charged particles. Correlations, which are significant, between used variables are ignored. Two likelihood functions have been built using actual data: one, $L_\pi$ to test the pion hypothesis\footnote{In
fact, in the case of the pion likelihood, two different functions
are built, one for $\pi^+$ and one for $\pi^-$, since the energy
deposition of the two particles is different, in particular 
at these energies. This is not necessary for electrons.
} (using $\pi^+\pi^-\pi^0$ events)
and the other, $L_e$ to test the electron hypothesis
(using Bhabha events). We require that candidate pion tracks have
a greater likelihood of being pions than of being electrons. We use as discriminating variable $\zeta=\log(L_\pi/L_e)$ and we accept as pions particles with $\zeta>0$, see fig.~\ref{likelihood}.
The efficiency for pions is \ab98\% and 3\% of the electrons survive rejection. We require that only one of the two particles satisfy the pion ID requirement (OR) achieving an efficiency of 100\% with some very small residual background. Requiring that both tracks are pions (AND) leads to a 5\% loss. We use the AND for evaluating backgrounds, see later.
\subsubsection{Kinematic Cuts}\label{sec:kincut} 
\begin{figure}[hb]
\begin{center}
\mbox{\epsfig{file=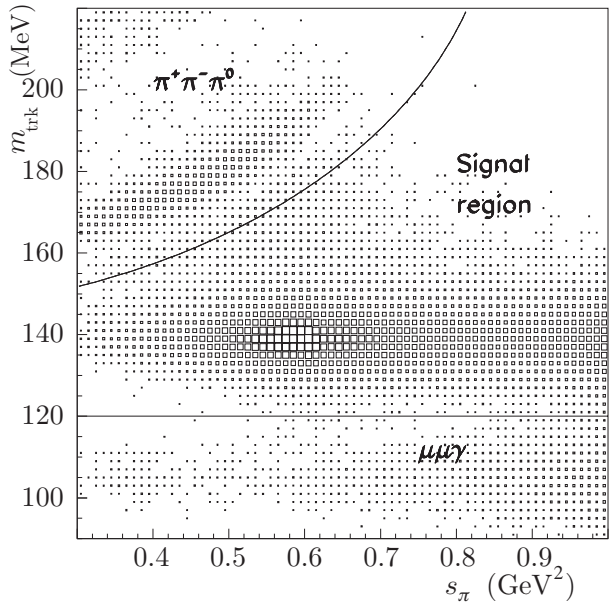,width=8cm}}
\caption{Kinematic separation, in the [$m_{\rm trk}, s_\pi$] plane, of signal and backgrounds after the $\zeta$ variable cut.}
\label{fig:mtrkvsq2}
\end{center}
\end{figure}

For an $x^+x^-\gamma$ final state Lorentz invariance allows us to compute the mass of $x$, which we call $m_{\rm trk}$ from
\begin{equation}
\left(M_\phi-\sqrt{{\bf p}_+^2+m^2_{\rm trk}}-\sqrt{{\bf p}_-^2+m^2_{\rm trk}}\right)^2-({\bf p}_++{\bf p}_-)^2=0
\label{eq:mtrk}
\end{equation}
For $\epm\to x^+x^-\gamma$, the  value of $m_{\rm trk}$ peaks at $m_\pi,\ m_\mu,\ m_e$ thus allowing selection of signal events.
The density distribution of the two track events in the [$s_\pi,\ m_{\rm trk}$] plane is very effective for separating signal from background. The distribution is shown in fig.~\ref{fig:mtrkvsq2} together
with the selection cut defined by: $(m_{\rm trk}>120)\odot(m_{\rm trk}<250-105\sqrt{1-s_\pi/0.85^2})\odot(m_{\rm trk}<220)$, all units in MeV. 
\pic\po\ events populate a curve in  the plane. The signal events above the region $m_{\rm trk}>140$ MeV are due to additional ISR or FSR radiation.  The cut on $m_{\rm trk}$ rejects a fraction of events with additional radiation. The implications are discussed in section 4.4.

\section{Event Analysis}
Data were taken in July through December 2001, for an integrated luminosity of \L=140.7 pb\up{-1}. After fiducial volume and selection cuts we find \ab\pt1.5,6, events.  Fig.~\ref{fig:nraw} gives the \pic\ mass squared distribution of the \pic\gam\ events in bins of 0.01 GeV\up{-2} in $s_\pi$.
The $\rho$ peak and the $\rho$-$\omega$ interference structure are clearly visible, demonstrating the excellent mass resolution of the KLOE detector.
To obtain the cross section, we subtract the residual 
background from this spectrum and divide by the selection
efficiency and the integrated luminosity:
\begin{figure}[ht]
\begin{center}
\mbox{\epsfig{file=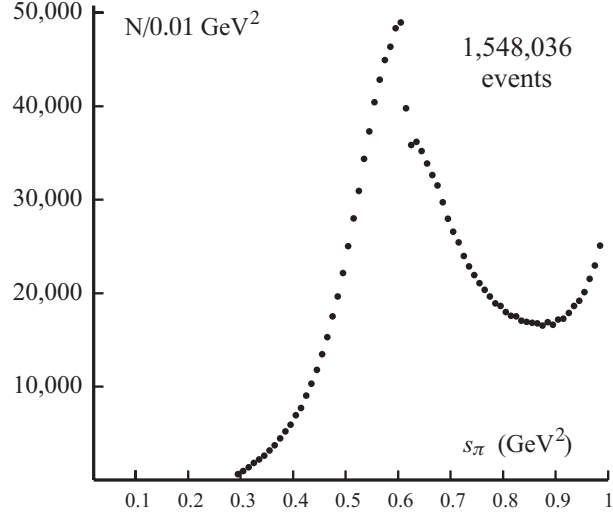,width=8cm}}
\caption{Events per 0.01 GeV$^2$ after fiducial volume and selection cuts.}
\label{fig:nraw}
\end{center}
\end{figure}
\begin{equation}
{\Delta\sig_{\pic\gam}\over\Delta s_\pi}=%
{\Delta N_{\pic\gam}\over\L\;\Delta s_\pi}=%
{\Delta N_{\rm obs}-\Delta N_{\rm bckgnd}\over\L\;\Delta s_\pi}\x%
{1\over\epsilon_{\rm cuts}}
\label{eq:sighad}
\end{equation}

\subsection{Background}
\begin{figure}[ht]
\begin{center}
\mbox{\epsfig{file=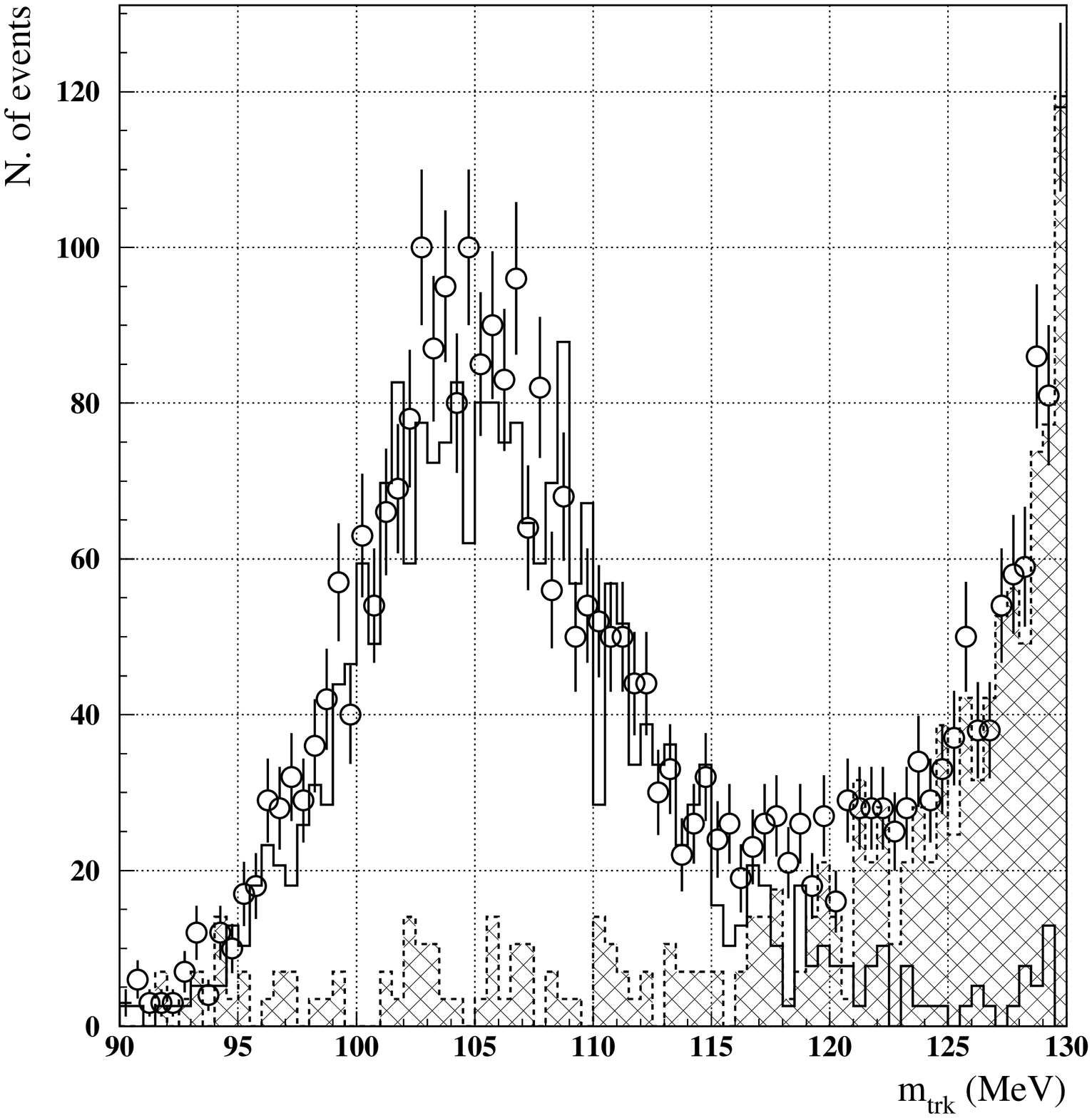,width=6.8cm}}
\mbox{\epsfig{file=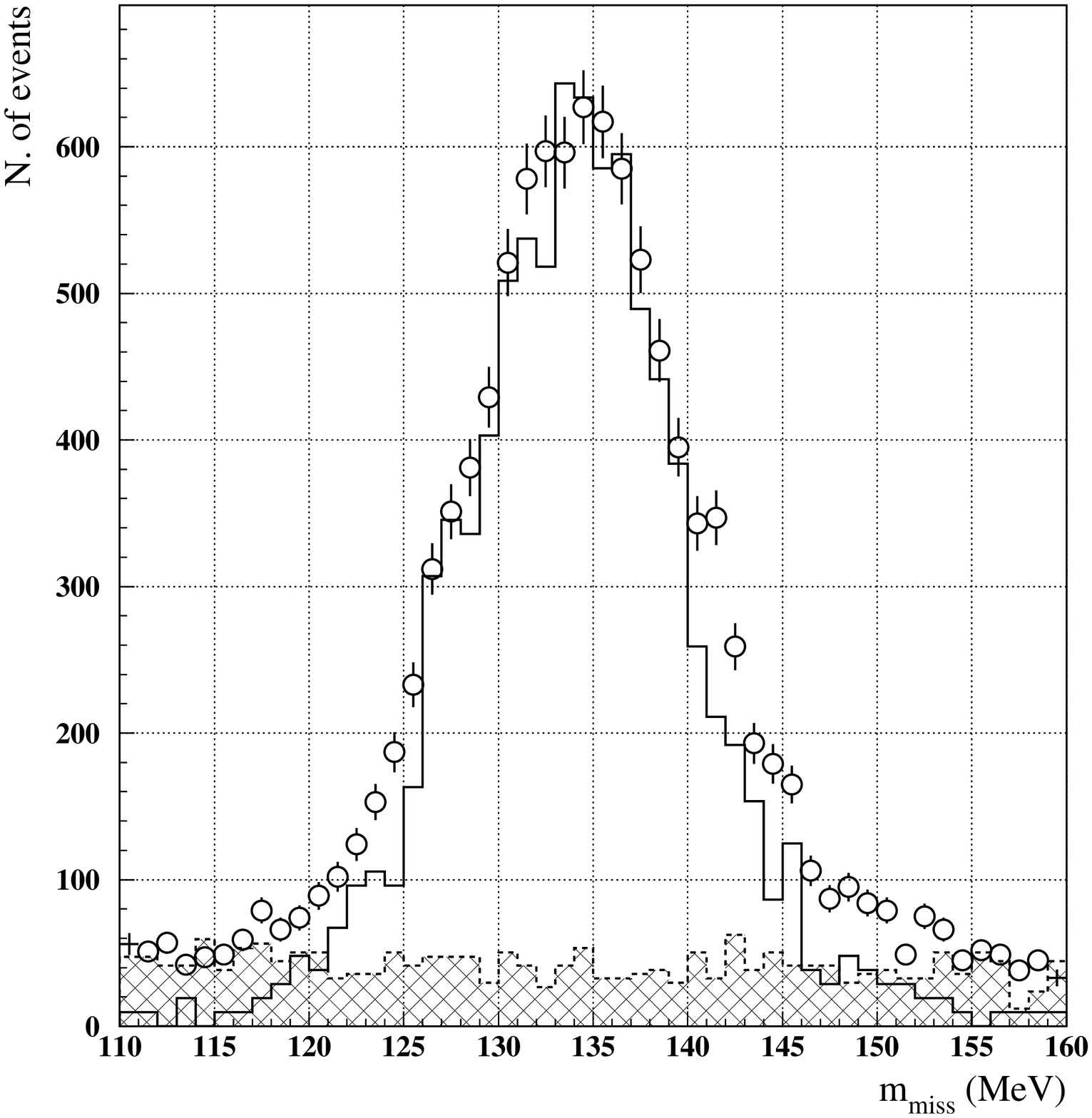,width=6.8cm}}
\caption{Muon mass fit, left and \po\ mass fit, right. These fits are used to estimate the \mmg\ and \pic\po\ backgrounds (shaded areas) to the \pic\gam\ channel.}
\label{fig:bkg1}
\end{center}
\end{figure}

\begin{figure}[ht]
\begin{center}
\mbox{\epsfig{file=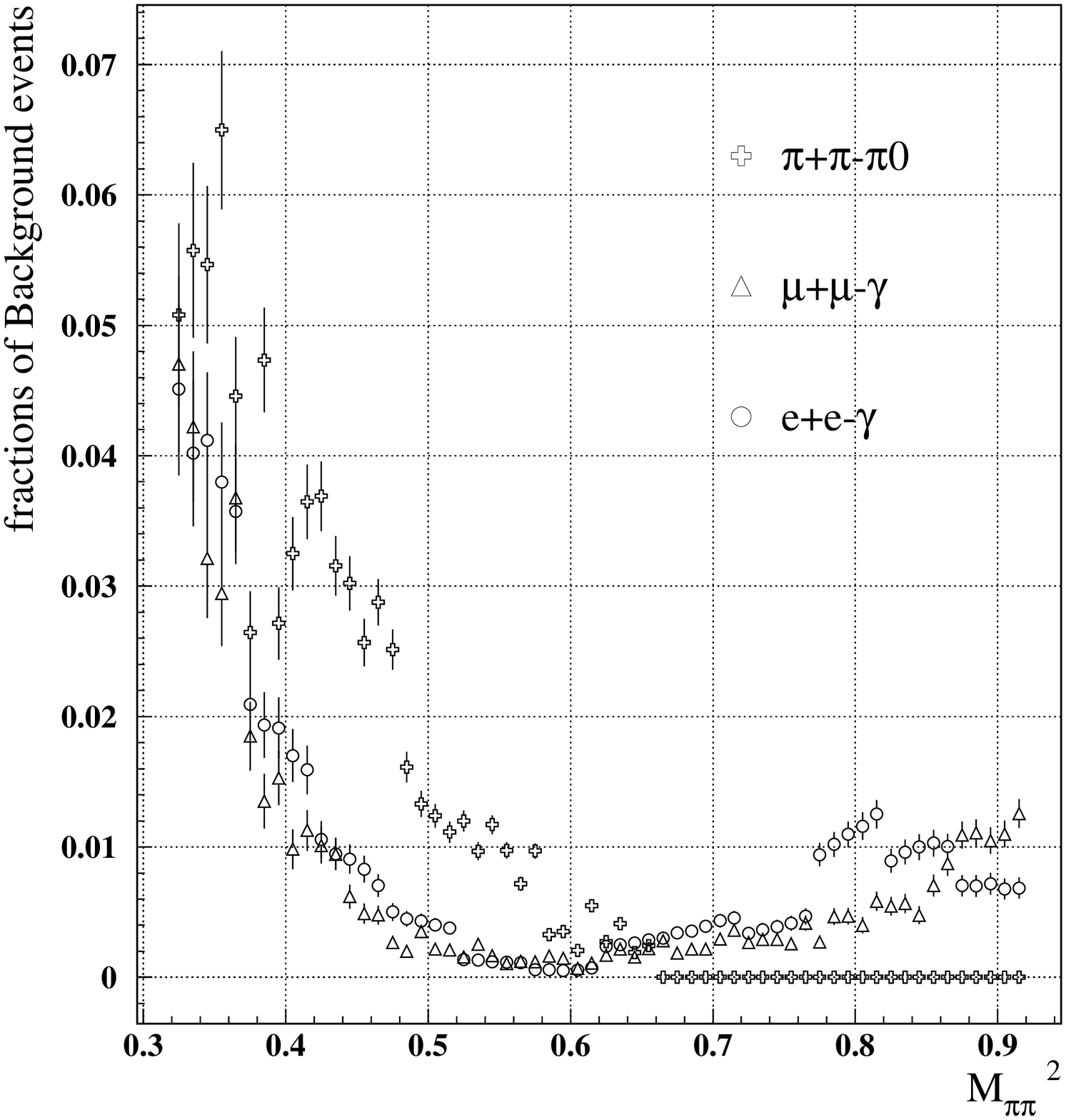,width=6.5cm}}
\mbox{\epsfig{file=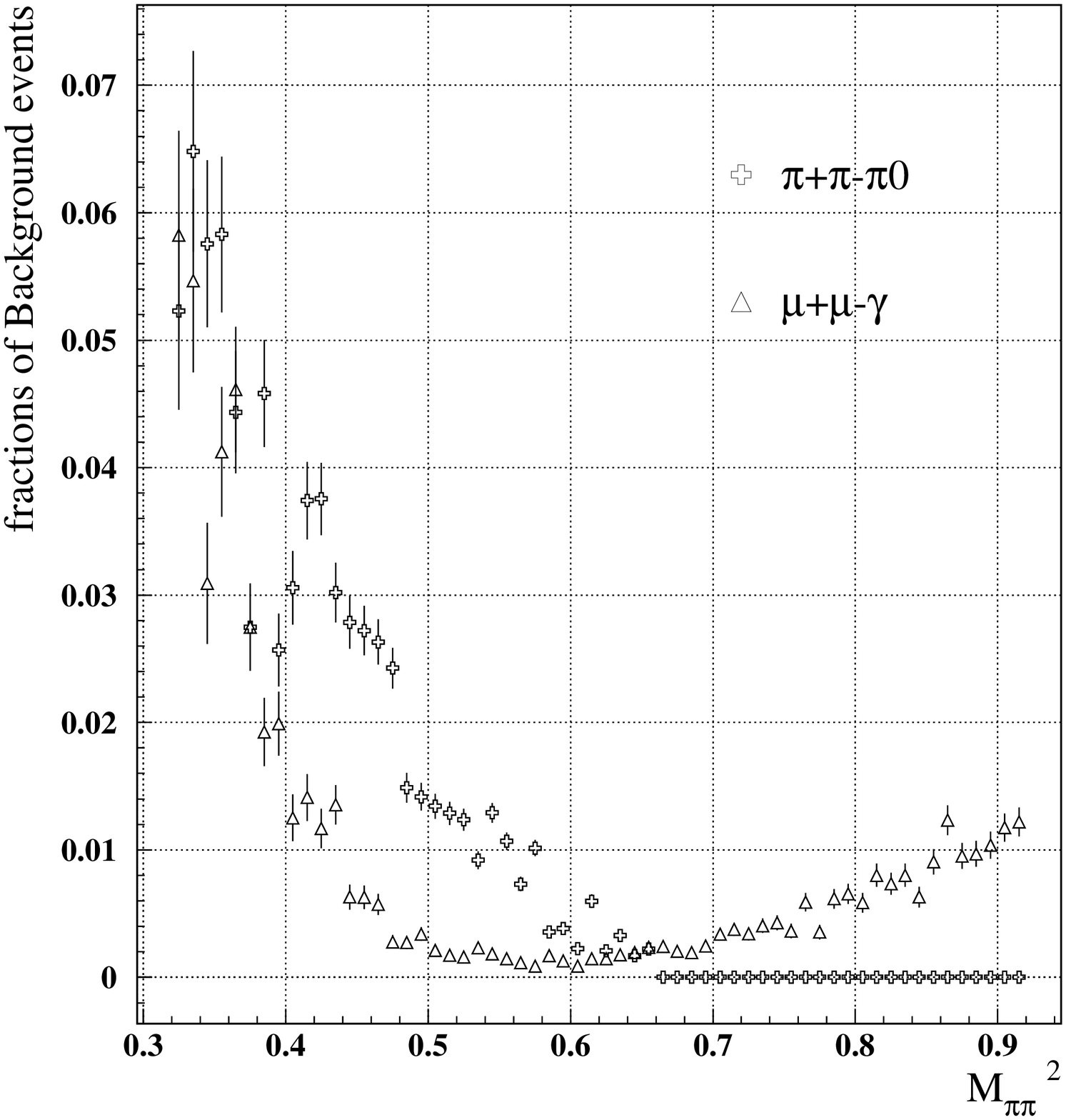,width=6.5cm}}
\caption{Left: backgrounds fractions requiring a single identified pion (OR) and right: two identified pions (AND) which removes completely the residual Bhabhas.}
\label{fig:bkg2}
\end{center}
\end{figure}

The [$s_\pi,\ m_{\rm trk}$] plane population of signal and background events is illustrated in fig.~\ref{fig:mtrkvsq2}.
The amount of background in the signal region is obtained by fitting the \mtrk\ spectrum of the selected events, in slices of $s_\pi$ with the \mtrk\ spectra for signal, \eeg\ and \mmg\ events obtained from Monte Carlo simulation. An example of such a fit is shown in fig.~\ref{fig:bkg1}, left.
Background from \ppp\ events appears at higher \mtrk\ values and the missing mass, $m^2_{miss} = (p_\phi - p_+ - p_-)^2$, peaks at $m(\po)$. The \ppp\ background count is obtained by fitting the $m_{\rm miss}$
distribution with the shapes obtained from the Monte Carlo simulation.
An example of such a fit is shown in fig.~\ref{fig:bkg1}, right.
The result of the fits is shown in fig.~\ref{fig:bkg2}
which shows the relative
contribution of the different backgrounds to the observed 
signal. The shape of the background distribution is well reproduced by the Monte Carlo simulation ensuring that systematic uncertainties are smaller than the fit errors.
Other possible contributions, such as the reaction \epm\to\epm\pic\ with the electrons going down the beam pipe, have been studied and shown to be negligible given our event selection criteria.

\subsection{Luminosity Measurement}

The  integrated luminosity is measured using
large angle Bhabha (LAB) events, with the KLOE detector 
itself. The effective Bhabha cross section 
at large angles 
($55^o < \theta_{+,-} < 125^o$) is about $430$ nb.  
This cross section is large enough so that the statistical error
on the luminosity measurement is negligible.
The number of LAB candidates, $N_{LAB}$, is counted and normalized to the 
effective Bhabha cross section, obtained by Monte Carlo: 
\begin{equation}
\int{\mathcal L}dt =
\frac{N_{LAB}(\theta_i)}{\sigma_{LAB}^{MC}(\theta_i)}
\cdot (1-\delta_{Bkg})
\end{equation}
The precision of this measurement depends on the correct inclusion of higher order terms in computing the Bhabha cross section. We use two independent Bhabha event generators: BHAGENF (\cite{berends}, \cite{drago}) and BABAYAGA (\cite{babayaga}). For each generator a systematic error of $0.5\%$ is quoted by the authors. The two result for the luminosity agree to better than 0.2\%.

\begin{figure}[ht]
\begin{center}
\mbox{\epsfig{file=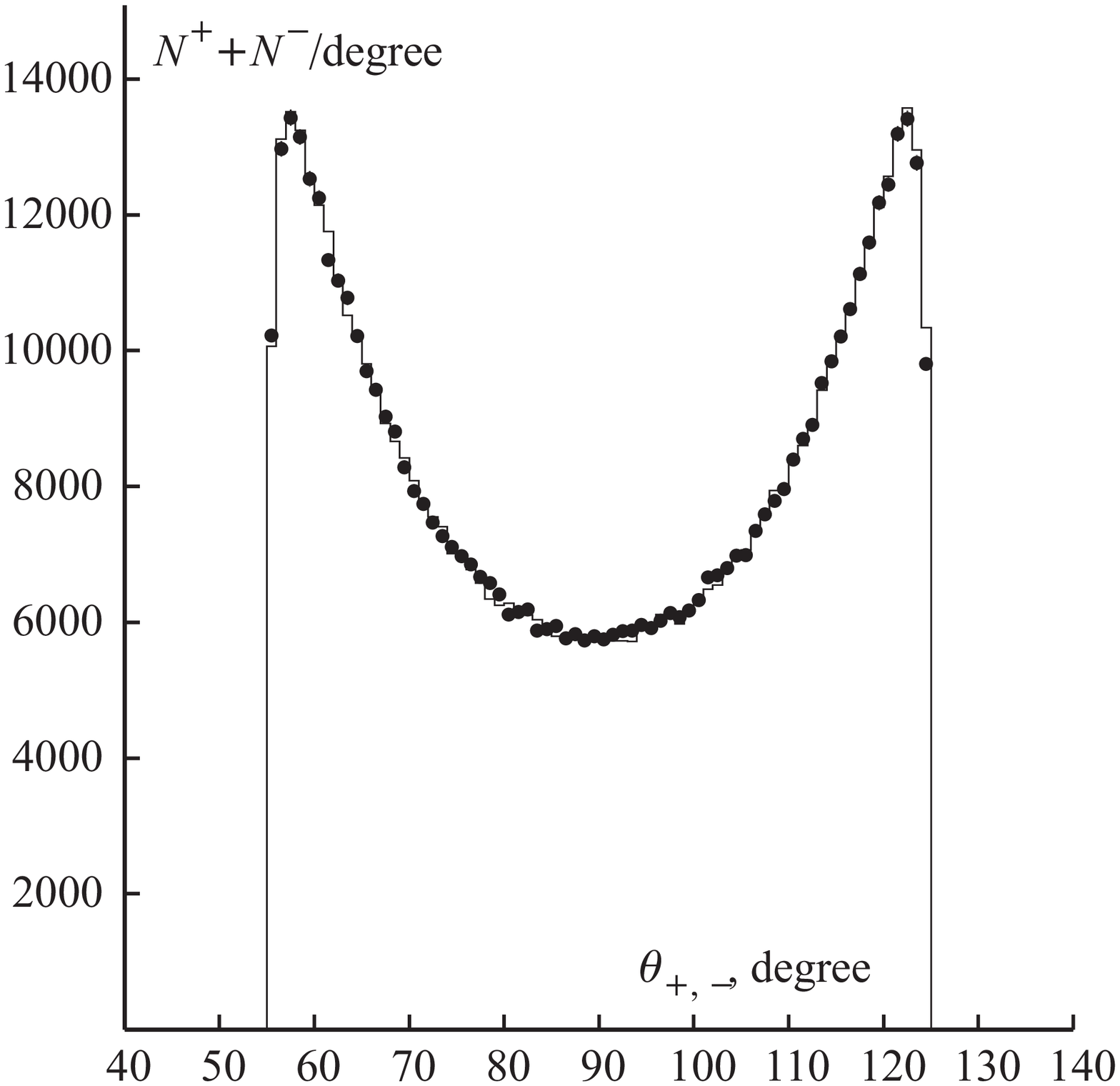,width=6.5cm}}
\mbox{\epsfig{file=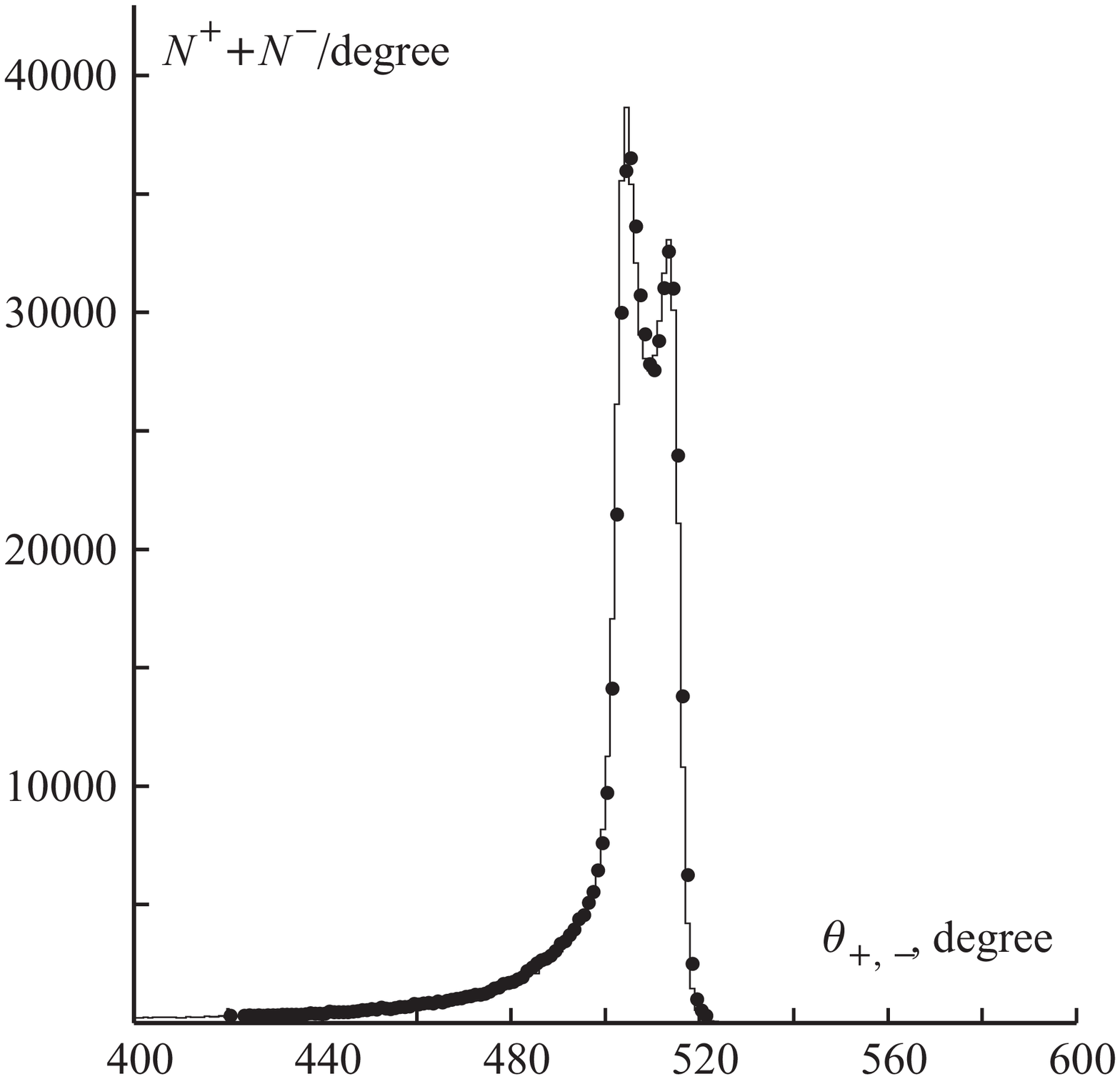,width=6.5cm}}
\caption{Comparison data-Monte Carlo
of the  $\theta_{+,-}$ (left) and $E_{+,-}$ (right) distributions
for the Bhabha events selected at large angle as described
in the text.}
\label{fig:bhabha}
\end{center}
\end{figure}

LAB events are selected with cuts on variables which are well simulated by the KLOE Monte Carlo simulation.
The electron and positron polar angle cuts, 55\deg$\,<\theta_{+,-}<\,$125\deg, is based on the calorimeter clusters, while the energy cuts, $E_{+,-}>400$ MeV, is based on drift chamber information. 
The background from $\mu^+ \mu^- (\gamma)$, $\pi^+ \pi^- (\gamma)$ and $\pi^+ \pi^- \pi^0$ events is well below $1\%$ and is subtracted. All selection efficiencies (trigger, EmC cluster, DC tracking) are $>99\%$ and are well reproduced by the detector simulation program. 
We also obtain excellent agreement between the experimental distributions ($\theta_{+,-}$, $E_{+,-}$) and those obtained from Monte Carlo simulation, see fig.~\ref{fig:bhabha}.
Finally, corrections are applied on a run-by-run basis for fluctuations in the center-of-mass energy of the machine and in the detector calibrations.
The experimental uncertainty in the acceptance due to all these effects is 0.4\%. We assign a total systematic error for the luminosity of 
$\delta\L=0.5\%_{\rm th}\oplus0.4\%_{\rm exp}$.
An independent check of the luminosity measurement is obtained using \epm\to\gam\gam\ events. We find  agreement to within $0.2\%$.

\subsection{Selection Efficiencies}
\subsubsection{Trigger, tracking and vertexing} 
These efficiencies have been obtained from control samples of \pic\po, \pic\ and \pic\gam\ events selected without requiring the vertex. The efficiencies have been parameterized as a function of single track parameters (charge, momentum and angle) combined using Monte Carlo simulation.
The largest source of systematic error in this procedure
arises from the difference
between the kinematics of the control samples and the
signal.
\subsubsection{Pion identification}. The efficiency for the $\zeta$ cut has been evaluated from data by accepting one track and studying the distribution of the other one. The systematic error introduced by the $\zeta$ cut is negligible, since we use the OR, \ie\ at least on particle is a pion, for which the efficiency is \ab100\% for \pic\gam\ events.
\subsubsection{Particle mass} The efficiency for the $m_{\rm trk}$ cut is obtained as a by-product of the residual background evaluation; the
result of the fit provides the efficiency in each $s_\pi$ bin.
However, this efficiency depends upon the way the Monte Carlo simulation treats multi-photon processes. The $m_{\rm trk}$ efficiency 
has been obtained with our reference Monte Carlo simulation, which uses the PHOKHARA version 1.0 generator.
PHOKHARA simulates the full NLO QED correction for \epm\ annihilation with the emission of additional photons 
from the initial state. In order to estimate the systematic
effect due to the generator we have repeated the exercise 
using a different generator, BABAYAGA ver.3.5, the same generator used 
in the luminosity measurement. In this generator, ISR is
treated using the parton shower approach. The resulting value for the
$m_{\rm trk}$ efficiency differs from that evaluated with
PHOKHARA by $0.2\%$. 
An additional contribution, which is treated separately, is the emission of both an ISR and FSR photon. 
This is discussed in the next section.
The contribution of the systematic error for each efficiency to the
total error  is shown in table~\ref{tab:syseff1}.

\begin{table}
\begin{center}
\begin{tabular}{||l|c||}
\hline
Acceptance & 0.3\% \\[-2mm]
Trigger & 0.6\% \\[-2mm]
Tracking & 0.3\% \\[-2mm]
Vertex & 1.0 \% \\[-2mm]
Likelihood & 0.1 \% \\[-2mm]
Track Mass & 0.2 \% \\[-2mm]
Background subtraction & 0.5 \% \\[-2mm]
Unfolding & 0.6 \% \\
\hline
Total exp systematics & 1.4 \% \\
\hline
\hline
luminosity & 0.6 \% \\[-2mm]
Vacuum Polarization &  0.2 \% \\
\hline
Total theor. systematics & 0.7 \% \\
\hline
\hline
FSR resummation & 2.0 \% \\
\hline
\hline
\end{tabular}
\caption{List of systematic uncertainties from the three sources: experimental, theoretical and ignoring FSR.}
\label{tab:syseff1}
\end{center}
\end{table}

\subsection{Unfolding the mass resolution}
\begin{figure}[ht]
\begin{center}
\mbox{\epsfig{file=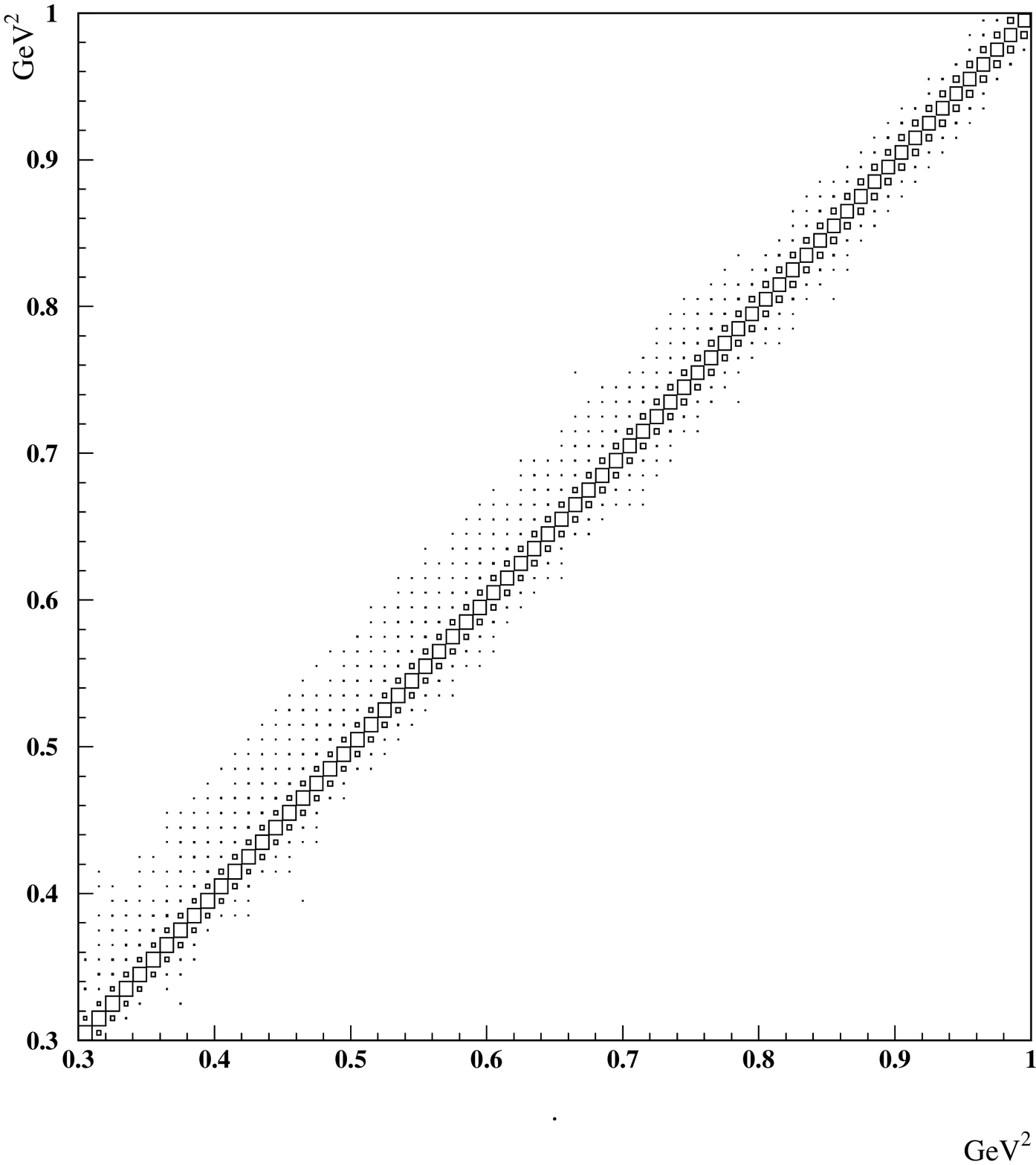,width=6.5cm}}
\mbox{\epsfig{file=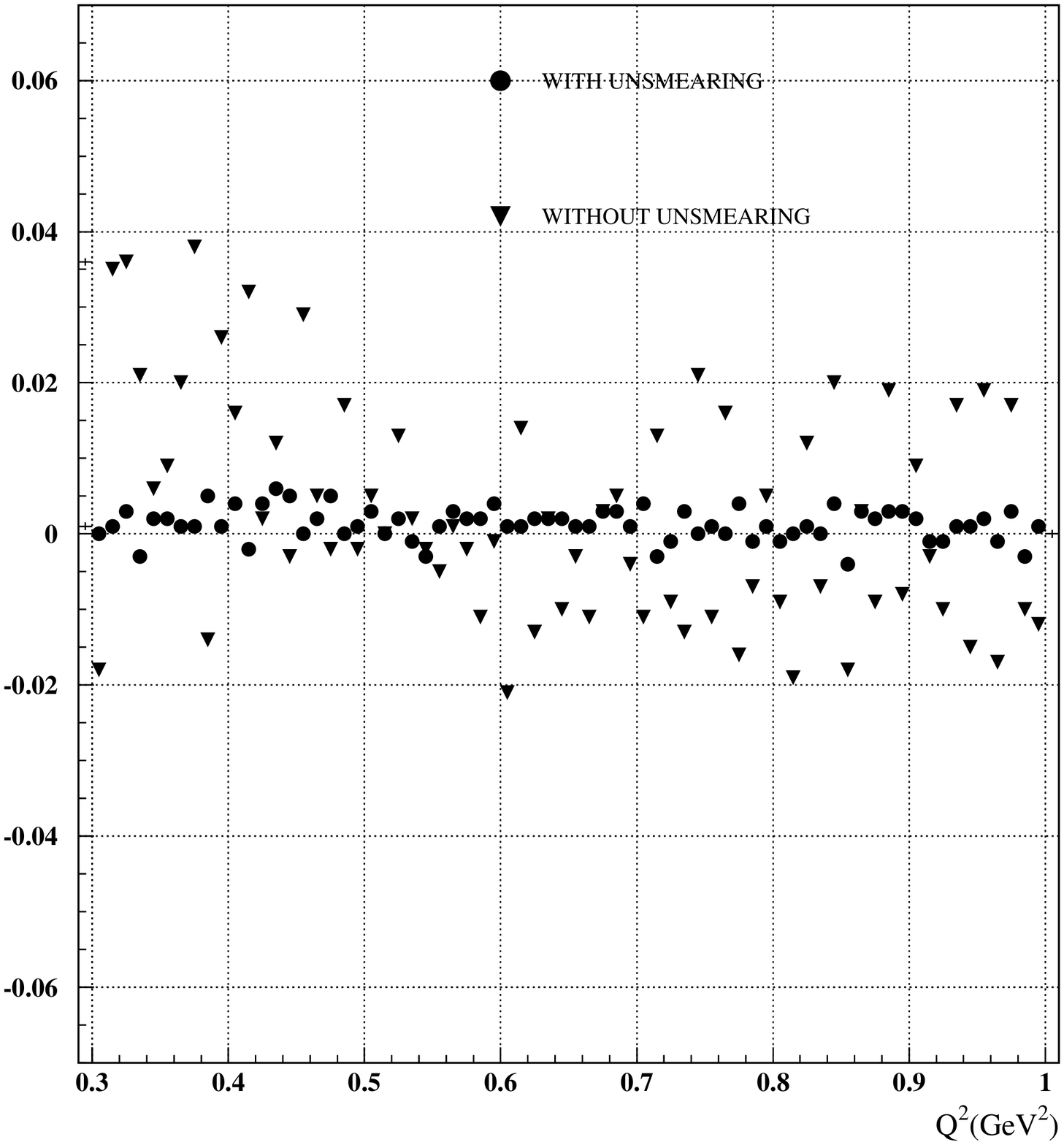,width=6.5cm}}
\caption{Left: smearing matrix representing the correlation
between generated and reconstructed $s_\pi$ value; the high precision
of the DC results in an almost diagonal matrix.
Right: fractional difference between reconstructed and 
generated spectrum with and without resolution unfolding.}
\label{fig:matrix}
\end{center}
\end{figure}

\def\minus{\hbox{$-$}}
In order to obtain $\dif\sigma_{\pi \pi \gamma}/\dif s_\pi$ as a function of the true $s_\pi$ value, we unfold the mass resolution from the measured $s_\pi$ distribution.
The measured value of $s_{\pi,{\rm obs}}$ is related to the true value by a resolution matrix {\bf G}($s_{\pi,{\rm true}}-s_{\pi,{\rm obs}}|s_{\pi,{\rm true}}$). This matrix has been generated by Monte Carlo simulation. 
The resolution matrix is nearly diagonal with $75\%$ of the events in the diagonal, $11\%$ in each of the first once-off-diagonal elements and only $3\%$ 
in the remaining elements. The matrix is graphically illustrated in fig.~\ref{fig:matrix}, left.
Unfolding is obtained by applying the inverse matrix {\bf G}$^{-1}$ to the data.
The result is shown in fig.~\ref{fig:matrix}, right, 
which shows the difference between 
the reconstructed and the generated spectra with and without the
application of the unfolding procedure.
This procedure does not affect the value of $a_\mu$, since the correlation is short range, $\simeq 97\%$ of the events being in the same or the adjacent bins.

\section{Results}

\begin{figure}[ht]
\mbox{\epsfig{file=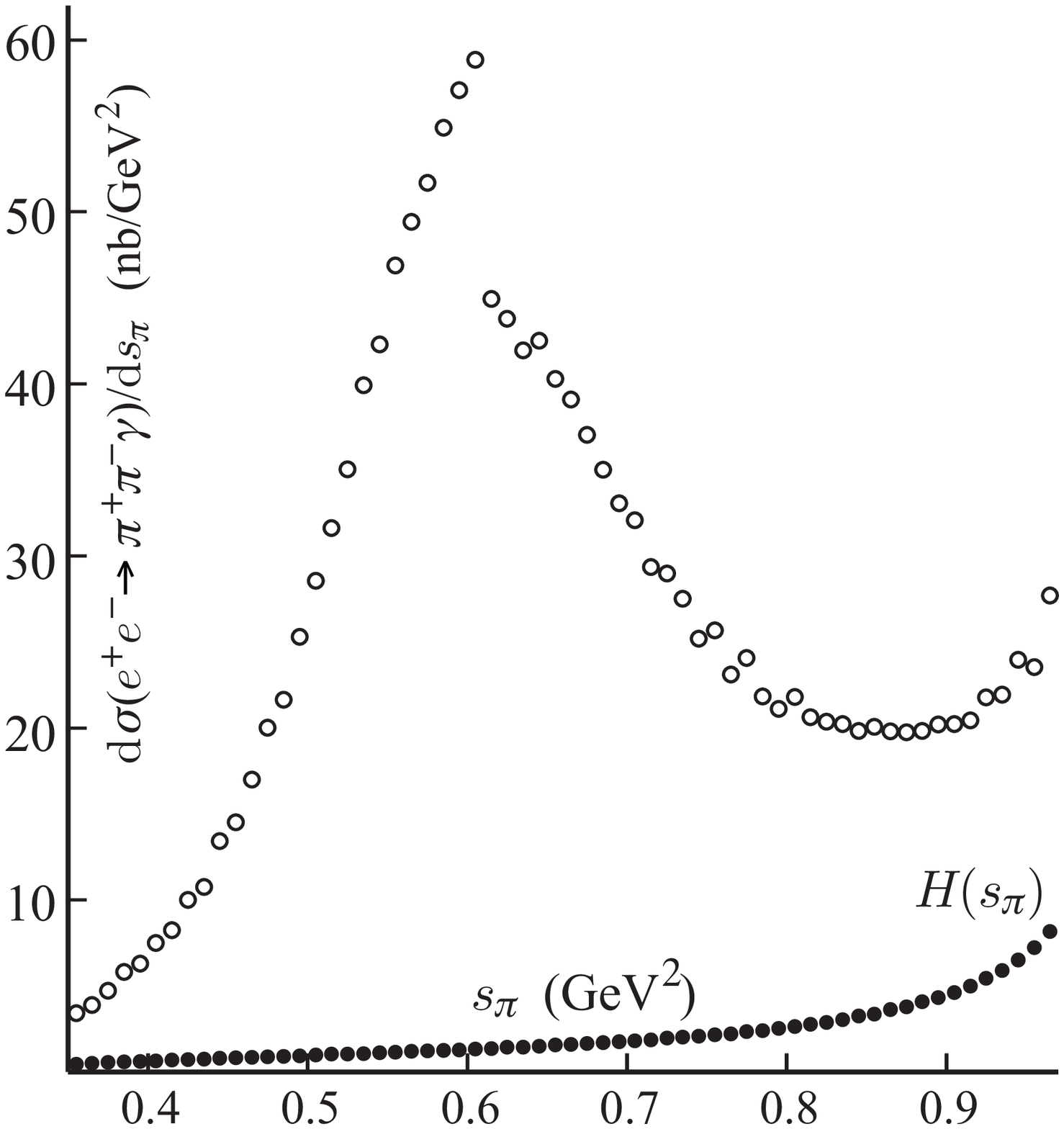,width=6.5cm}}\kern5mm
\mbox{\epsfig{file=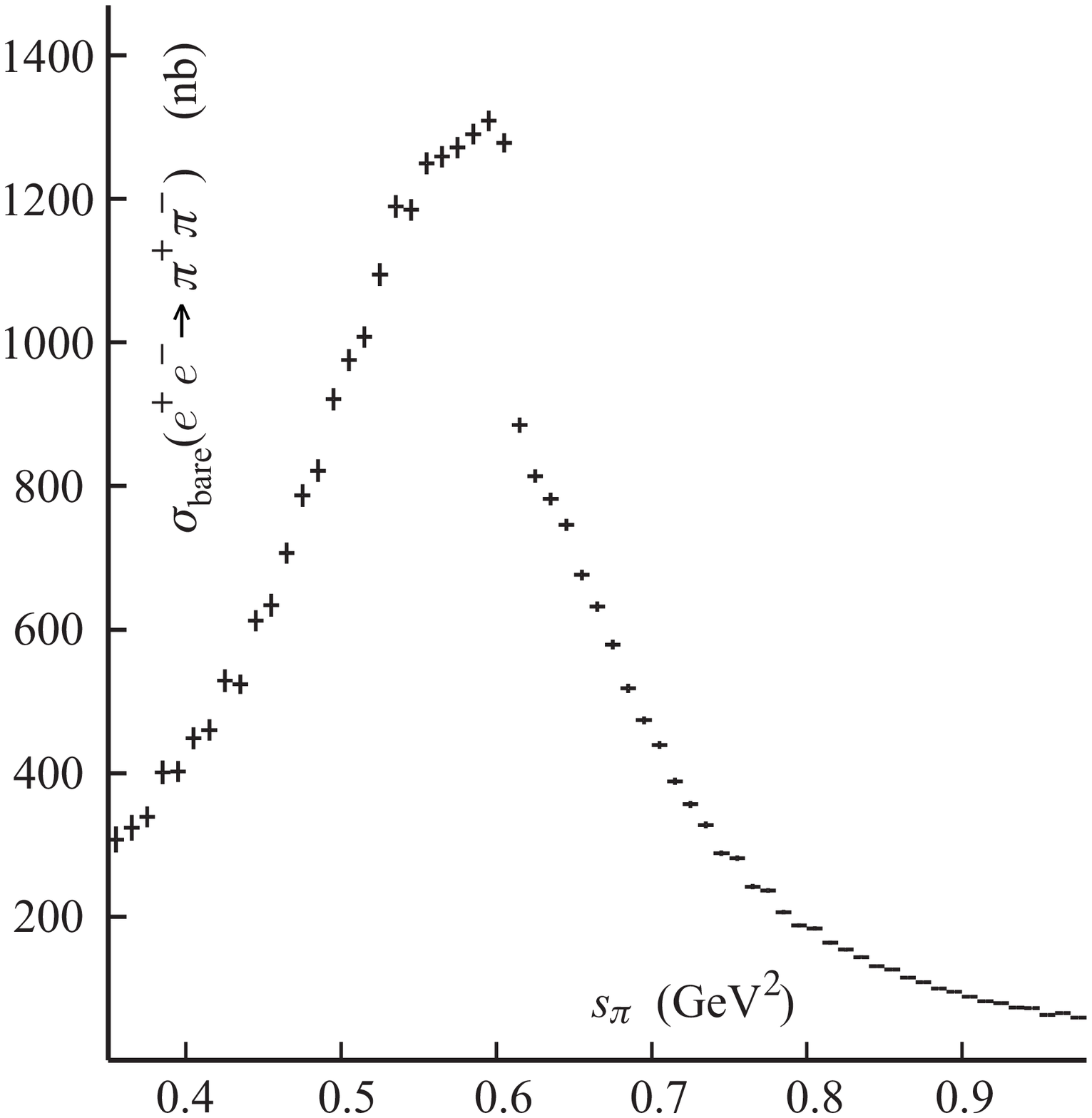,width=6.5cm}}
\caption{Left. Cross section for $e^+e^-\rightarrow \pi^+\pi^-
\gamma$. The radiator $H(s_\pi,\overline\theta_\gamma)$, is also shown.
Right. {\it Bare} cross section for $e^+e^-\rightarrow \pi^+\pi^-$.}
\label{fig:sigppg}
\end{figure}

The cross section for  $e^+ e^- \rightarrow \pi^+ \pi^- \gamma$, after applying the corrections described above, is shown in fig.~\ref{fig:sigppg}. 
In order to get the $e^+ e^- \rightarrow \pi^+ \pi^- $ cross
section, the radiator function $H(s_\pi,\overline\theta_\gamma)$ is needed. 
The radiator is obtained from PHOKHARA, setting $F_\pi(s_\pi)=1$ and {\it switching off} the vacuum polarization of the intermediate photon
in the generator. 
The radiator $H(s_\pi,\overline\theta_\gamma=15\deg)$ is also shown in fig.\ref{fig:sigppg}, left.

\subsection{FSR and Vacuum Polarization corrections}

Part of the multiphoton events are removed by the $m_{trk}$ cut. The effect of this cut is evaluated using PHOKHARA 1.0 which, however, does not include soft radiation in the final state accompanying one hard photon from the initial state. 
An preliminary estimate done with PHOKHARA-II, shows that this effect is smaller than 2\%. 
We {\it do not correct} here for FSR photons and we include in the theoretical systematic error the full value (2\%) due to this effect.
In order to obtain the pion form factor, or the {\it bare }
cross section, vacuum polarization effects must be subtracted. 
This can be done by correcting the cross section for the running of
$\alpha$, see refs. \cite{hoefer}, \cite{davhoe}, as follows:
\begin{equation}
\sigma_{\rm bare} = \sigma_{\rm dressed} \left( \frac{\alpha(0)}{\alpha(s)}
\right)^2
\end{equation}
We have used an approximate, but quite adequate for the purpose estimate of $\alpha(s)$\cite{vacpol}. Fig. \ref{fig:sigppg}, right, shows the resulting cross section.

\subsection{Results and comparisons with existing data}
We have used the {\it bare} cross section to evaluate of $\delta a_\mu^{\rm had}$ 
in the region covered by the CMD-2 experiment ($0.37<s_\pi<0.93$).
The resulting value (in 10$^{-10}$ units) is :
\begin{equation}
\delta a_\mu^{\rm had}=374.1\pm1.1_{\rm stat}\pm 5.2_{\rm syst}\pm 2.6_{\rm theo}\left.\; ^{+7.5}_{-0.0}\right|_{\rm FSR}
\end{equation}
to be compared with:
\begin{equation}
\delta a_\mu^{\rm had}\hbox{(CMD-2)}=368.1\pm2.6_{\rm stat}\pm 2.2_{\rm syst+theo}
\end{equation}

The statistical error is negligible. 
The systematic error is at the moment at the $1.4\%$ value,
but there is still room to improve it down to $\simeq
1\%$.
Clearly the dominating error is the one 
we have conservatively assigned to the 
FSR radiation effect. This error  will be substantially reduced 
 once the new PHOKHARA version will be inserted into the KLOE MC.

The central value of our result in the region $0.37<s_\pi<0.93$ is 
slightly larger than the one obtained by CMD-2. 
The discrepancy is, at the moment, not significant (0.5 $\sigma$).
This difference is, however, not equally distributed in $s_\pi$, as summarized below.\footnote{This is our evaluation of $a_\mu^{\rm had}\hbox{(CMD-2)}$, based on the values tabulated in ref.~\cite{cmd2}}
$$\vbox{\halign{\hfil#\hfil\qquad&\hfil # \hfil\qquad&\hfil#\hfil\cr
$s_\pi$, GeV\up2&$\delta a_\mu$,~KLOE&$\delta a_\mu$,~CMD-2\cr
0.37 -- 0.60&$256.2\pm4.1(^{+5.1}_{-0}\ {\rm FSR})$&$249.7\pm2.2$\cr
0.60 -- 0.93&$117.9\pm2.1(^{+2.3}_{-0}\ {\rm FSR})$&$119.8\pm1.1$\cr}}$$
Our data differ from the CMD-2 results mostly below the $\rho$-peak.
However, for the mass squared range $0.6<s_\pi<0.9$, our data confirm the discrepancy between $\tau$ data and \epm\to\pic\ results, which is \ab10-15\%.

\section*{Acknowledgements}
We thank the DA$\Phi$NE team for their efforts in maintaining low background running 
conditions and their collaboration during all data-taking. 
We want to thank our technical staff: 
G.~F.~Fortugno for his dedicated work to ensure an efficient operation of 
the KLOE Computing Center; M.~Anelli for his continous support to the gas system and the safety of the detector; A.~Balla, M.~Gatta, G.~Corradi and G.~Papalino for the maintenance of the electronics; M.~Santoni, G.~Paoluzzi and R.~Rosellini for the general support to the detector; 
C.~Pinto (Bari), C.~Pinto (Lecce), C.~Piscitelli and A.~Rossi for their help during shutdown periods.
This work was supported in part by DOE grant DE-FG-02-97ER41027; by EURODAPHNE, contract FMRX-CT98-0169; by the German Federal Ministry of Education and Research (BMBF) contract 06-KA-957; by Graduiertenkolleg `H.E. Phys. and Part. Astrophys.' of Deutsche Forschungsgemeinschaft, Contract No. GK 742; by INTAS, contracts 96-624, 99-37; and by TARI, contract HPRI-CT-1999-00088.

\end{document}